\documentstyle[prl,multicol,aps,epsf]{revtex}
\begin{document}
\title{Quantum Phase Pumping}
\author{Fei Zhou}
\address{Physics Department, Princeton University, Princeton, NJ 08544}
\maketitle

\begin{abstract}
In this Letter we consider the adiabatic charge transport through 
a normal mesoscopic sample 
sandwiched by superconductors without modulation of local chemical
potentials. 
The deformation of coherent quasiparticles 
in the normal metal
in the presence of periodically changing phases in
the superconductors leads to a
charge transport. 
Both the magnitude and the phase dependence of 
the charge transferred per period through the sample  
strongly depend on temperature.
\end{abstract}

\begin{multicols}{2}

Recently adiabatic charge transport in mesoscopic samples 
generates much interest both experimentally
and theoretically\cite{Marcus,Spivak,Brower,Zhou98}. 
In the presence of slowly and periodically changing external 
fields, the spatial electron density profile follows 
periodical modulation of local {\em chemical
potentials}. 
In the presence of two a.c. charging gates, $\{g_1(t), g_2(t)\}$,
this leads to a d.c. adiabatic
charge transport across the sample in the {\em classical limit}.
The amount of charge transport equals the amount of "flux" of
a topological field $\pi_{12}({\bf g})$ threading the area
enclosed by the trajectory ${\cal C}$ swept by ${\bf g}=\{g_1(t), g_2(t)\}$
in a 2 dimensional parameter space ${\cal M}_2$.
At low temperature when the dephasing length is
longer than the sample size, electrons remain coherent.
$\pi_{12}({\bf g})$, unlike in the classical limit where it is a
constant field, develops very rich structures and is a sample specific,
random function of ${\bf g}$ in ${\cal M}_2$ space.  
Due to quantum interference effects,
the adiabatic charge transport at low temperature therefore is
of a random sign and magnitude\cite{Marcus,Brower,Zhou98}.

It still remains a question how the many body correlation  
enriches the structure of the topological field 
$\pi_{12}({\bf g})$ and affects the adiabatic charge transport.
In this letter, we exam an example 
where a mesoscopic sample is in contact with superconductors.
At $T \ll \Delta$(the superconductor energy gap), 
electron-like excitations in these samples  
are reflected into hole-like excitations through Andreev
reflections at superconductor-metal(SN) interfaces.  
Quasiparticles inside the normal
metal at low temperatures are superpositions of electrons
and holes, with zero effective charges\cite{Golubov,Atland,Fraham,Zhou95}.
For a superconductor-normal metal-superconductor
(SNS) junction, 
the resultant excitation spectrum in the metal has little
resemblance to that of a normal Fermi liquid: 
An energy gap of many body nature is opened up at Fermi energy, which 
remains finite while the single particle level spacing goes to zero
\cite{Golubov,Atland,Fraham,Zhou95}. 
The transport coefficients are also renormalized.

When a voltage is applied across the junction, the phase difference is 
changing according to Josephson relationship. For the geometry shown in
Fig.1, the phase differences of two SNS junctions are given as  
\begin{equation} 
\chi_{1,2}(t)=\frac{2eV t}{\hbar}+\delta_{1,2}
\end{equation}
with $\delta_{1,2}$ as the phase differences when $V=0$.
Electrons experiencing Andreev reflections would acquire
different phases at SN interfaces. Thus, 
electron-hole superpositions of quasiparticles 
depend on phase differences of superconductors.
The energy gap in the quasiparticle spectrum 
decreases from a finite value to zero when the phase difference
between two superconductors is changing from zero to
$\pi$\cite{Atland,Fraham,Zhou95}.  
Quasiparticles of energy $\epsilon$ at time t
can be represented by $(\nu_1(\epsilon, t), \nu_2(\epsilon, t))$,
with $\nu_0\nu_1(\epsilon, t)$ as the average density of states
in region 1, $\nu_0\nu_2(\epsilon, t)$ in region 2. (The density of states in
the normal metal outside region 1,2 equals its bulk value $\nu_0$, 
doesn't depend on $\chi_{1,2}(t)$.) 
As time evolves, $\chi_{1,2}$ changes and 
$\nu_{1,2}(\epsilon, t)$ oscillates
with a frequency $2eV/\hbar$. 
When $\chi_{1(2)}$ is far away from from  $\pi$,  
quasiparticles far below the energy gap extend only outside
region 1(2). When $\chi_{1(2)}$ approaches $\pi$, the gap diminishes; 
the low energy quasiparticles start to extend into region 1(2).   
This leads to a continuous 
deformation of quasiparticles of energy $\epsilon$, characterized
by the motion of $(\nu_1(\epsilon, t), \nu_2(\epsilon, t))$(see Fig.2).
It is worth emphasising that unlike the situations in
\cite{Marcus,Brower,Zhou98}, the local chemical
potential in this case remains constant as a function of time and the 
deformation of many body wave functions 
is possible only when the phase rigidity is present\cite{kin}.
For this reason, we call the phenomena discussed here as 
{\bf phase pumping}.

In the following,
we show that deformation of coherent quasiparticle
wave packets via phase differences 
eventually leads to a d.c. current in this case. 
The adiabatic charge transport in the presence of 
the changing phase differences of superconductors
is determined by {\em the sensitivities of
the diffusion constants and the
density of states to boundary conditions}.
At low temperatures, the topological field
is localized in the parameter space, leading to a very singular
phase dependence of d.c. charge transport. 
The charge transport
approaches zero as a power law function of temperature
in both high temperature and low temperature limits. 

To calculate the charge transport in this limit
we introduce Keldysh technique generalized to
Nambu space\cite{Keldysh,Larkin77},

\begin{equation}
i\hat{\tau}_z \frac{\partial}{\partial t}\hat{G}
+i\frac{\partial}{\partial t}\hat{G} \hat{\tau}_z
+[ H, \hat{G}] + I_{col}=0
\end{equation}
where $H$ is the Hamiltonian of electrons,
including random impurity potentials.
$I_{col}$ is the electron-phonon collision
integral.
$\hat{G}({\bf r}, {\bf r'}, t, t')$ is Keldysh
Green function matrix defined in Nambu 
space

\begin{equation}
\hat{G}=\left(
\begin{array}{cc}
\tilde{G}^R, \tilde{G}^K \\
0, \tilde{G}^A
\end{array}
\right), 
\tilde{G}^P=\left(
\begin{array}{cc}
G^P, iF^P \\
-iF^P, -G^P 
\end{array}
\right),
\end{equation}
$P=R, A, K$ and
$\hat{\tau}_z=\left(
\begin{array}{cc}
\tau_z, 0\\
0, \tau_z
\end{array}
\right)$,
$\tau_z$ is z-component of Pauli matrix in Nambu space.
Let $\tilde{G}^K_{\bf p}(\epsilon, {\bf r}, t)$ be Keldysh
component of the semiclassical Green function defined as

\begin{eqnarray}
\hat{G}_{\bf p}(\epsilon, {\bf r}, t)
=\frac{i}{\pi}\int d\xi_{\bf p} d{\bf r'} dt' \exp(
i\epsilon t'-i{\bf p} \cdot {\bf r'})
\nonumber \\
\hat{G}({\bf r}+{\bf r'}/2, {\bf r}-{\bf r'}/2, t+t'/2, t-t'/2)
\end{eqnarray}
where $\xi_{\bf p}={\bf p}^2/2m -\epsilon_F$, $\epsilon_F$
is the fermi energy.
In diffusion limit,
$\hat{G}_{\bf p}=\hat{G}(\epsilon, {\bf r}, t) +{\bf n}\cdot l\nabla
\hat{G}(\epsilon, {\bf r}, t)$, ${\bf n}$ is the unit
vector along ${\bf p}$. 
In the adiabatic approximation, the solution of
Eq.2 for Keldysh
component is written in term of two
nonequilibrium functions $f_{1}, f_2$ and
$\tilde{G}^{R,A}$,

\begin{eqnarray}
\tilde{G}^K(\epsilon, {\bf, r})=
(\tilde{G}^R(\epsilon, {\bf r})-\tilde{G}^A(\epsilon, {\bf r})) 
(1-2n_F(\epsilon)) \nonumber \\
+\tilde{G}^R(\epsilon, {\bf r})(f_2+\tau_z f_1) -
(f_2 +\tau_z f_1) \tilde{G}^A(\epsilon, {\bf r}), 
\end{eqnarray}
with $n_F$ as Fermi distribution function,
$f_{1(2)}(\epsilon)$ is an even(odd) function of $\epsilon$.

Following Eqs.2,5, in the first order adiabatic approximation,
the equations of $f_{1,2}$ and $G^R$ take forms of

\begin{eqnarray}
&&D\nabla \cdot D_2(\epsilon,{\bf r})\nabla f_2(\epsilon, {\bf r}) =
\frac{\partial n_F}{\partial \epsilon}
\int^{\epsilon}_0 d\epsilon' \frac{\partial \nu(\epsilon, {\bf r})}
{\partial t}, \nonumber \\
&&\nabla \cdot  D_1(\epsilon, {\bf r})\nabla f_1(\epsilon, {\bf r})=0
\nonumber \\
&&i\epsilon [\tau_z, \tilde{G}^R(\epsilon, {\bf r})]
+ D\nabla \cdot ( \tilde{G}^R(\epsilon, {\bf r}) \nabla 
\tilde{G}^R(\epsilon, {\bf r}))=0,
\end{eqnarray}
where $D$ is the diffusion constant in bulk metals;
$\nu(\epsilon, {\bf r}) =Tr(\tilde{G}^R\tau_z -\tau_z \tilde{G}^A)$,
$D_1(\epsilon, {\bf r})=Tr(1-\tilde{G}^R\tau_z\tilde{G}^A\tau_z)$,
$D_2(\epsilon, {\bf r})= Tr(1-\tilde{G}^R\tilde{G}^A)$, 
$Tr$ represents trace in Nambu space. 
Here we neglect $f_1$'s contribution in first equation in the
leading order of $T/\epsilon_F$.
To derive the right hand side of the first equation
in Eq.6, we use an identity
\begin{equation}
\frac{\partial H}{\partial t}\nu(\epsilon, {\bf r}) 
=\int^\epsilon_0 d\epsilon 
\frac{\partial}{\partial t}\nu(\epsilon, {\bf r})
+\frac{1}{2}\frac{\partial}{\partial t}
\int^{+\infty}_{-\infty} d\epsilon 
\nu(\epsilon, {\bf r})
\end{equation}
which is valid when the external parameters vary little over the 
characteristic length scale of $G^{R, A}$.
The second term in the right hand side of Eq.7 vanishes 
following the sum rule. Inside the leads, quasiparticles 
remain in equilibrium and $f_1(\epsilon, {\bf r})=
f_2(\epsilon, {\bf r})=0$ at the lead-metal boundaries. 
At SN boundaries, $G^R=0$, $F^R=\exp(i\chi)$ at $\epsilon
\ll \Delta$.
Electron-phonon collisions are neglected in Eq. 6 in the low temperature
limit when the diffusion relaxation dominates. 

Finally,
in the leading order of ${1}/k_Fl$, the current can be expressed as

\begin{eqnarray}
&&{\bf J}=
D\nu_0 \int d\epsilon [\frac{\epsilon}{2\epsilon_F}D_2(\epsilon, {\bf r})
\nabla f_2(\epsilon, {\bf r})
+ D_1(\epsilon, {\bf r}) 
\nabla f_1(\epsilon, {\bf r})], 
\nonumber \\
\end{eqnarray}
with $\nu_0$ the density of states 
in bulk metals.

\vspace{-0.1cm}
\begin{figure}
\begin{center}
\epsfbox{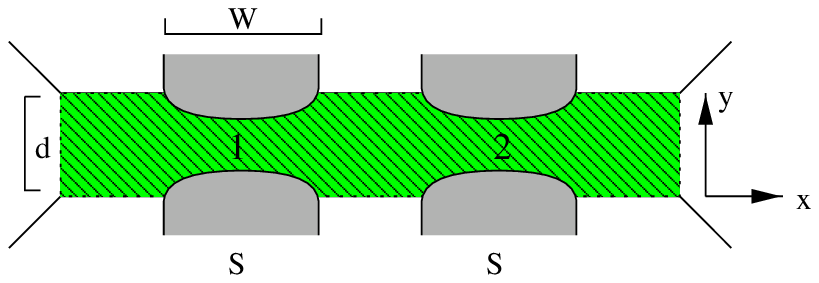}
\leavevmode
\end{center}
\vspace{-0.7cm}
Fig.1. Geometry of the sample, $W \gg d$. 
Shaded bars stand for superconductors; the patterned region
represents a disordered normal metal. 
\end{figure}

For the quasi $1-D$ geometry shown in Fig.1 where 
$W, L\gg d\gg \sqrt{D/\Delta}$, 
$G^{R, A}(x,y)$ of junction 1,2 can be 
approximated
with $G^{R,A}(y)$ calculated in the leading order
of $(d/W)^2$($L$ is the length of the sample along x-direction) 
at given $\chi_{1,2}$. 
This allows us to integrate Eq. 6 along y-direction independently.
The resultant equations 
$f_1, f_2$ and $G^{R, A}$ have only x-dependences.
From Eq.6 and its boundary conditions, we find $f_1=0$ 
and  

\begin{eqnarray}
&&f_2=\frac{\partial n_F(\epsilon)}{\partial\epsilon}
\int^x_0 dx'
\frac{1}{D{D}_2(\epsilon, x')}
[\frac{\partial {\cal X}(\epsilon, x')}{\partial t}
\nonumber \\
&&-\int^L_0 dx' \frac{1}{{D}_2(\epsilon, x')}
\frac{\partial {\cal X}(\epsilon, x')}{\partial t}
(\int^L_0 dx' \frac{1}{{D}_2(\epsilon, x')})^{-1}].
\end{eqnarray}
Here we introduce
${\cal X}(\epsilon, x) =\int^x_0 dx'\int^\epsilon_0 d\epsilon' 
\nu(\epsilon', x')$,
and ${\rho}^{-1}(\epsilon, x)={{D}_2(\epsilon, x)}\int^L_0 dx' 
{D}^{-1}_2(\epsilon, x')$.


We want to emphasis that
$G^{R, A}$ depends on time only through boundary conditions $\chi_{1,2}$,
$\chi_{1,2} \in [-\infty, +\infty]$.
Wave functions are periodical in such a $2D$ space, 
and $G^{R, A}(\chi_1, \chi_2)=G^{R,A}
(\chi_1+2\pi, \chi_2+2\pi)$. Furthermore,
time reversal symmetry of the ground state at $\chi=0$ requires 
that physical quantities
like $D_{1,2}$ or $\nu$ are even functions of $\chi$.  
Evidently, it is more convenient
to introduce a compact $2D$ space ${\cal M}_2$ of $g_{\alpha}$, 
$g_{\alpha}=\cos\chi_\alpha$,
$\alpha=1,2$. $g_{\alpha} \in [-1, 1]$.

Following Eq. 6, in the adiabatic approximation, the nonequilibrium 
distribution function depends on time {\em locally} through
$\chi_{1,2}(t)$, $\partial_t\chi_{1,2}(t)$ or ${\bf g}$, $\partial_t {\bf 
g}$. Correspondingly, the time dependence of
$\rho(\epsilon, x)$, ${\cal X}(\epsilon, x)$ defined
after Eq. 9 can be expressed 
through their dependence on ${\bf g}$.
Thus, the time derivative of ${\cal X}$ is represented by a 
one-form defined in such a 2D parameter space.  
Substituting Eq.9 back into Eq. 8 and using Stoke's theory to
carry out the integral of 
the one-form along a close trajectory ${\cal C}$ in ${\cal M}_2$\cite{Zhou98},
we obtain the charge transport along $x$ direction per period
in term of a two-form defined 
in ${\cal M}_2$

\begin{eqnarray}
&&Q=\int_{\cal S}
\pi_{\alpha\beta} dg_\alpha\wedge dg_\beta \nonumber \\
&&\pi_{\alpha\beta}=
\nu_0\int d\epsilon \frac{\epsilon}{\epsilon_F}
\frac{\partial n_F}{\partial\epsilon} 
[\tilde{\pi}_{\alpha\beta}(\epsilon, {\bf g})-
\tilde{\pi}_{\beta\alpha}(\epsilon, {\bf g})]
\nonumber \\
&& \tilde{\pi}_{\alpha\beta}(\epsilon, {\bf g})=
\int^L_0 dx  
\frac{\partial \rho(\epsilon, x)}{\partial g_\alpha} 
\frac{\partial {\cal X}(\epsilon, x)}{\partial g_\beta} 
\end{eqnarray}
where ${\cal S}$ is the region enclosed by trajectory
${\cal C}$ in ${\cal M}_2$.
In the leading order of $(d/W)^2$, 
${\pi}_{\alpha\beta}$ only depends on the 
sensitivities of $D_2, \nu$ with respect to 
${\bf g}$, which 
reflects {\em the degree of deformation
of coherent wave packets} at point ${\bf g}$ in ${\cal M}_2$.

\begin{eqnarray}
&& \tilde{\pi}_{\alpha\beta}(\epsilon, {\bf g})=
\frac{\partial {\Sigma}(\epsilon, g_\beta)}{\partial g_\beta} 
\frac{\partial {\Xi}_\beta(\epsilon, {\bf g})}{\partial g_\alpha}. 
\nonumber \\
&&{\Sigma}(\epsilon, g_\alpha)=
\int^\epsilon_0 d\epsilon' \nu(\epsilon', g_\alpha),
\nonumber \\
&&\Xi_\beta(\epsilon, {\bf g})=\eta_0
[\sum_\gamma \frac{1}{D_2(\epsilon, g_\gamma)}\eta_0
+(1-2\eta_0)]^{-1} \times \nonumber \\
&&[ 1 - \eta_\beta+ \eta_0 
\sum_\gamma
(\frac{1}{2}\delta_{\beta\gamma}+{\theta}(x_\gamma -x_\beta))
(\frac{1}{D_2(\epsilon, g_\gamma)}-1)]
\end{eqnarray}
with $D_2, \nu$ 
calculated at given $g_\alpha$ in the limit $W \gg d$.
$\eta_0=W/L$, $\eta_\beta={(x_\beta +W/2)}/{L}$,
$x_\beta$ is the x-coordinate of the center of $\beta$ region. 
${\theta}(x)$ is a step function.
{\em In the absence of phase rigidities, 
$\Sigma, \Xi$ do not
depend on ${\bf g}$ and $\pi_{\alpha\beta}=0$.}
$Q$ naturally represents {\em a connection  
over a tangent fiber bundle $TM$}, where the fiber is
one-form $\Sigma(\partial\Xi/\partial g) dg$ embedded in 
the tangent manifold of ${\cal M}_2$ and the base manifold is 
${\cal M}_2$. Such a geometric point of view was 
first 
emphasized in \cite{Thouless,Avron} for isolated quantum mechanical
systems and generalized to statistical systems in \cite{Zhou98}, 
where kinetic processes are important. 
Generally speaking, $Q$ is not proportional to
the area enclosed by trajectory ${\cal C}$ in the base manifold
${\cal M}_2$ because 
$\pi_{\alpha\beta}$ depends on ${\bf g}$.
{\em Absence of the area law, which usually 
holds in a classical system, signifies strong correlation effects.}

Following the normalization condition
we rewrite $G^R=\cos\theta(\epsilon),
F^R=\exp(i\chi(\epsilon))\sin\theta(\epsilon)$,
where $\theta=\theta_1+i\theta_2$, which is  
obtained as a solution to the last equation
in Eq. 6\cite{Zhou95}. 
$D_1(\epsilon)=\cosh^2\theta_1(\epsilon)$, 
$D_2(\epsilon)=\cos^2\theta_2(\epsilon)$
and $\nu(\epsilon)=\cos\theta_1(\epsilon)\cosh\theta_2(\epsilon)$.
It is important to notice
that $\theta_1(\epsilon)\neq \pi/2$
or $D_2, \nu \neq 0$  only when $\epsilon 
\geq E_g$, where $E_g$ is a function of $\chi$\cite{Zhou95}

\begin{equation}
E_g(\chi \leq \pi)=E_c\left \{ \begin{array}{cc}
C_2(1-C_1\chi^2),  \mbox{$\chi \ll \pi$;} \\
C_3(\pi -\chi), \mbox{$\pi-\chi \ll \pi$.}
\end{array}\right.
\end{equation} 
Here $E_c=D/d^2$, $C_1=0.91, C_2=3.122, C_3=2.43$.
Following Eq. 12, the density of states has a gap, which 
closes only when $\chi=\pi$.
At energies much larger than $E_c$, the phase dependence of
$\theta_{1,2}$ 
decays exponentially as a function of energy,

\begin{equation}
\theta_{1, 2}=\exp(-\sqrt{\frac{\epsilon}{E_c}})(1+g_{1,2}). 
\end{equation}
Taking into account these results, we find
\begin{eqnarray}
\tilde{\pi}_{\alpha\beta}(\epsilon, {\bf g})
=\left\{ \begin{array}{cc}
\eta_0^2 e^{-\sqrt{{\epsilon}/{E_c}}}
[\theta(\eta_\alpha-\eta_\beta)+\eta_\beta -1], 
\mbox{$\epsilon \gg E_c$;} \\
F(g_\alpha+1, g_\beta+1, \epsilon/E_c), \mbox{$\epsilon \ll E_c$.} 
\end{array}
\right.
\nonumber \\
\end{eqnarray}

\vspace{-0.2cm}
\begin{figure}
\begin{center}
\epsfbox{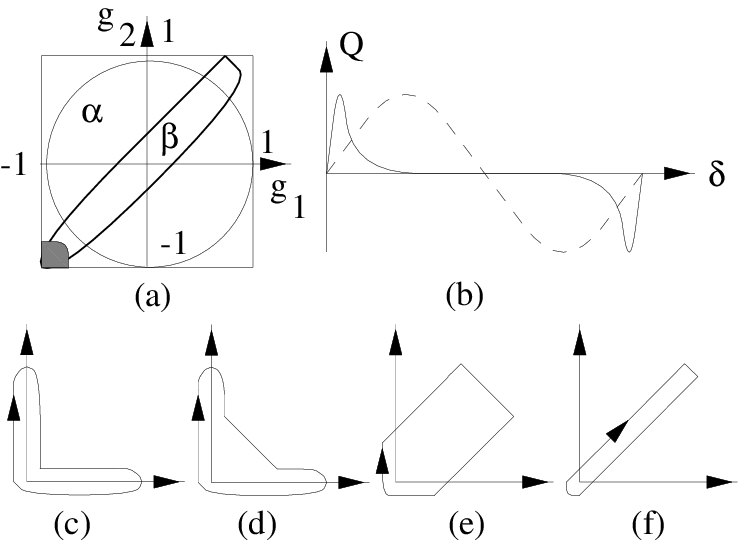}
\leavevmode
\end{center}
\vspace{-0.5cm}
Fig.2. a). Distribution of $\pi_{\alpha\beta}$ in ${\cal M}_2$
at $kT \ll E_c$. $\pi_{\alpha\beta}$ is nonzero only in the shaded region
of size $kT/E_c$. Trajectory $\alpha$ represents the case $\delta=\pi/2$
and trajectory $\beta$ for $\delta\sim kT/E_c$.
b). $Q(T)$ as a function of $\delta$ at $kT \gg E_c$ (dashed line)
and $kT \ll E_c$(solid line). 
c), d), e), f) are
plots for $(\nu_1(T, t), \nu_2(T, t))$ at different $\delta$ 
and at $kT \ll E_c$.
c). $\delta \gg kT/E_c$;  d). $\delta$ approaches $kT/E_c$; 
e). $\delta \sim kT/E_c$; f). $\delta \sim 0$.
\end{figure}

$F$ is zero when $|g_{\alpha, \beta} +1| \gg (\epsilon/E_c)^2$
and approaches unity when $|g_{\alpha, \beta}+1| \sim
(\epsilon/E_c)^2$. 
Eqs. 11, 14 indicate that $\tilde{\pi}_{\alpha\beta}(\epsilon, {\bf g})$
is exponentially small and is a simple sinusoidal function
of $\chi$ at $\epsilon \gg E_c$.
In the opposite limit, the phase dependence of $\tilde{\pi}_{\alpha\beta}$ 
depends strongly upon the high harmonics.

At high temperatures, 
the main contribution to $\pi_{\alpha\beta}$ 
is from electrons of energies of order $E_c \ll kT$ since 
contributions of electrons with energies of order $kT$
are exponentially small. 
Substituting Eq.14 into Eq. 10,
we obtain the charge transport at high temperature due to
deformation of quasiparticles, 
\begin{equation}
Q(T)=N \left(\frac{W}{L}\right)^2
\left(\frac{E_c}{\epsilon_F}\right)^2\frac{E_c}{kT}F_1(\delta).
\end{equation}
$N$ is the number of electrons inside the sample.
$F_1(\delta)$ is a smooth function of $\delta$, as shown
in Fig.2b(the dashed line); $F_1(\delta=0, \pi)=0$.
The small factor $E_c/\epsilon_F$ is from the
asymmetry of the electron-hole spectrums at Fermi energy;
$E_c/kT$ originates from the fact that only wave packets of this small
fraction of quasi-particles can be deformed. 
When $kT$ is larger than the Josephson coupling energy,
the thermal phase slippage takes place and $Q(T)=0$.

At low temperature, $\pi_{\alpha\beta}$
is determined by $\epsilon \sim kT \ll E_c$.
Following Eq. 14, it is nonzero only in a small region
where $|g_{\alpha, \beta}+1|$ is smaller than $kT/E_c(\ll 1)$,
as shown in Fig2.a. 
At zero temperature, $\pi_{\alpha\beta}$ becomes completely
localized at $g_\alpha=g_\beta=-1$ and equals zero in the rest
of ${\cal M}_2$ plane.
The localization of $\pi_{\alpha\beta}$ in ${\cal M}_2$
indicates very different phase dependences of charge transport
in the low temperature and high temperature 
limits. Consider the case $\delta=\pi/2$. The corresponding
trajectory $\alpha$ in Fig.2a does not contain the small shaded region
where $\pi_{\alpha\beta}\neq 0$. Following Eq.10, $Q$ equals zero
in this case.
To have nonzero charge transport, trajectory
${\cal C}$ is required to enclose a small portion in ${\cal M}_2$
of radius $kT/E_c$
where $\pi_{\alpha\beta} \neq 0$, and $\delta$ needs 
to be close to $kT/E_c$. 

Physically, charge can be
transferred only when the density of states
at energy $kT$ in regions 1,2 (see Fig. 1)
is changing simultaneously. 
It can be demonstrated more explicitly by directly looking
at deformation of wave packets,   
which can be characterized by
the trajectory of the vector $(\nu_1(T, t), \nu_2(T, t))$
in a 2-D space, as shown in Fig.2. c). d). e). f).
Here $\nu_0\nu_{1,2}(T, t)$ are the density of states
of energy $kT$ in region 1,2  at time $t$. 
According to Eq. 12, 
the energy gap $E_g$ in the quasiparticle spectrum
closes only at $\chi=\pi$ and the density of states
at low energy remains zero for most of the period unless
$\chi$ is in the vicinity of $\pi$.
When $\delta \gg kT/E_c$, $\nu_1$ changes while $\nu_2$ remains zero
and vice versa. The trajectory therefore repeats itself when the time
evolves and the area enclosed is zero. 
When $\delta$ approaches $kT/E_c$, $\nu_2$ becomes nonzero while
$\nu_1$ is changing. The trajectory starts to enclose 
a finite area. The area reaches a maximum when $\delta \sim kT/E_c$.
When $\delta$ decreases further towards zero, $\nu_1$ and 
$\nu_2$ start to change in phase; the trajectory becomes
a narrow strip with its width proportional to $\delta$.
As $\delta=0$, the trajectory again repeats itself in the 2D space
and the area enclosed is zero.
The trajectory of $\delta < 0$ and that of $\delta >0$
obeys mirror symmetry, i.e. the trajectory of $\delta <0$ can be obtained
by reversing the direction of the trajectory of $-\delta$. 
As a consequency, 

\begin{equation}
Q(T)= N \left(\frac{W}{L}\frac{kT}{\epsilon_F}\right)^2 
F_2(\delta, \frac{kT}{E_c}), 
\end{equation}

where

\begin{equation}
F_2(\delta, \delta_0)=sign\delta \left\{
\begin{array}{cc}
|\delta| \delta_0, \mbox{$|\delta| \ll \delta_0$;} \\
\delta_0^2, \mbox{$|\delta| \sim \delta_0$;} \\
\exp(-|\delta|/\delta_0), \mbox{$|\delta| \gg \delta_0$.}
\end{array}\right.
\end{equation}
The contribution to the charge transport in the limit
$|\delta| \gg \delta_0$ is from the quasiparticles
of energy of order $\delta \times E_c$, population of which
is much smaller than that of typical excitations of energy $kT$.
Though there is an energy gap in the spectrum, charge transport is
not quantized in this case because $kT$ can be 
higher than $E_g(\chi)$ when $\chi\rightarrow 0$.

From Eqs. 15, 16, we find the charge transport 
vanishes as the temperature goes to zero because of
the electron-hole symmetry and decreases as $1/T$ at high temperature
limit due to the suppression of electron-hole coherence. 
$Q$ as a function of temperature
develops a maximum at $kT \sim E_c$. 
At asymptotically low temperature, one can no longer
neglect the finite dwell time of the electrons in  
region 1,2. $Q(T)$ saturates at the value
$N (d/W)^2(E_c/\epsilon_F)^2$ when $kT \sim E_c (d/W)^2$.
Another type of quantum interference effect, i.e.
mesoscopic fluctuations is negligible in the present
case because the dimension along $z$ direction
in Fig.1. is assumed to be infinite.

In conclusion, we find that electrons can be transferred
out of dots with the help of the phase rigidity of coherent 
wave packets. In the presence of small amount magnetic 
impurities, 
the gap of quasi particle spectrums closes as far as 
phase differences are in the vicinity of $\pi$, the size of which is
determined by the pair breaking rate. In this case, 
$\pi_{\alpha\beta}$ would be delocalized
in ${\cal M}_2$ space and the pronounced peak of charge transport
as a function of $\delta$ as shown in Fig.2b will be smeared out. 
When the depairing rate is further increased, 
coherence effects become exponentially small and $Q(T)\approx 0$.

Finally, in the metallic limit studied in this letter, 
quantum phase slippage effects are negligible.

The author acknowledges stimulating discussions with 
B. Altshuler, C. Marcus, B. Spivak. 
He is grateful to Andrew Yeh for carefully reading the manuscript.
This work is supported by ARO under DAAG 55-98-1-0270.
He also likes to thank ICTP, Trieste, Italy and        
NECI, Princeton, USA for their hospitalities.

\end{multicols}


\begin{thebibliography}{99}
\bibitem{kin}
From a kinetic point of view,
such deformation creates nonequilibrium
distribution functions of electrons.
Especially, when the density of states at
a given energy increases(decreases), electrons at the corresponding levels
become underpopulated(overpopulated) and the effective temperature of the
electronic system is lower (higher) than the thermal bath temperature. 
At low temperatures, imbalance charges accumulated 
due to the asymmetry of the density of states at Fermi energy relax via 
diffusing out of the sample.
\end{thebibliography}
\end{document}